\documentclass[11pt]{article}
\usepackage[margin=1in]{geometry}
\usepackage{amsmath}
\usepackage{amssymb}
\usepackage{graphicx}
\usepackage{bm}
\usepackage{xcolor}
\usepackage{placeins}
\usepackage[style=numeric-comp,giveninits=true,doi=false,url=false,isbn=false,
eprint=true,sorting=none,backend=biber,maxbibnames=99]{biblatex}
\usepackage{hyperref}
\hypersetup{colorlinks=true,citecolor=blue,linkcolor=blue,urlcolor=blue}
\renewbibmacro{in:}{\ifentrytype{article}{}{\printtext{in\space}}}
\DeclareFieldFormat[article]{title}{#1}
\DeclareFieldFormat[article]{pages}{#1}
\DeclareFieldFormat{eprint:arxiv}{arXiv\addcolon\space #1}
\AtEveryBibitem{\clearfield{month}\clearfield{number}\clearfield{issue}\clearfield{issn}\clearfield{doi}\clearfield{url}}

\newcommand{\runin}[1]{\par\medskip\noindent\textit{#1}—}

\begin{document}
\begin{center}
{\Large Topological Origin of Directional Amplification in Arbitrary-Dimensional Non-Hermitian Lattices\par}
\vspace{0.5em}
{\normalsize Zhenxiang Wu\textsuperscript{1}, Linfeng Zhang\textsuperscript{1,2,3}, and Haitan Xu\textsuperscript{1,2,3,*}\par}
\vspace{0.35em}
\begin{minipage}{0.95\textwidth}
\centering\small
\textsuperscript{1}National Laboratory of Solid State Microstructures, School of Advanced Manufacturing Engineering, Nanjing University, Suzhou 215163, China\\
\textsuperscript{2}Jiangsu Key Laboratory of Quantum Information Science and Technology, Nanjing University, Suzhou 215163, China\\
\textsuperscript{3}Shishan Laboratory, Nanjing University, Suzhou 215163, China\\[0.25em]
\textsuperscript{*}Corresponding author: \texttt{haitanxu@nju.edu.cn}
\end{minipage}
\end{center}
\vspace{0.6em}
\begin{abstract}
In non-Hermitian lattices, combinations of gain, loss and nonreciprocal couplings can generate exponentially growing directional amplification with applications in signal processing and precision measurement.
    An intriguing question is whether such amplification has a universal topological origin in arbitrary dimensions.
    Here we establish a dimension-independent framework to uncover the topological origin of directional amplification based on the concept of amoeba, which is the logarithmic image of the singularities of the response generating function.
    The complement of the amoeba in the logarithmic parameter space consists of connected components, each carrying a distinct and discrete topological label.
    We show that the asymptotic response is governed by the component extending toward negative infinity.
    When the origin of the logarithmic parameter space is outside this component and not on its boundary, there must exist some directions in which the response is exponentially amplified. Otherwise, there is no directional amplification in any direction. Our results provide a general topological criterion for classifying the responses in arbitrary-dimensional non-Hermitian lattices.
\end{abstract}
\vspace{1em}

\runin{Introduction}Non-Hermiticity can fundamentally reshape both the spectrum and the response of a physical system~\cite{Ashida02072020,barzanjeh_non-reciprocity_2026}, giving rise to exceptional points~\cite{xu_topological_2016,el-ganainy_non-hermitian_2018,doi:10.1126/science.aap9859,doi:10.1126/science.aar7709,PhysRevLett.123.066405,PhysRevLett.127.186601,doi:10.1126/sciadv.adl5037}, non-Hermitian skin effect~\cite{PhysRevLett.121.086803,PhysRevB.102.205118,zhang_observation_2021,zhang_universal_2022,wztw-l8wg,PhysRevLett.134.176601,zhao_two-dimensional_2025,zhang_harmonic_2026} and directional amplification~\cite{PhysRevLett.120.087401,wen_unidirectional_2022,PhysRevB.106.024301,PhysRevA.107.053515,gr8d-b74y}.
The interplay between non-Hermiticity and topology has led to a richer variety of topological phases~\cite{PhysRevX.8.031079,PhysRevX.9.041015,PhysRevLett.124.086801,RevModPhys.93.015005}, with a bulk-boundary correspondence that is different from the conventional Bloch paradigm~\cite{PhysRevLett.121.086803,PhysRevLett.121.136802,PhysRevLett.123.066404,helbig_generalized_2020,PhysRevX.14.021011}.
Besides the spectrum, this interplay also reshapes the linear response of a lattice~\cite{PhysRevLett.122.143901,PhysRevLett.125.180403,parto_enhanced_2025}.
The connection between directional amplification and topology in one dimension has attracted broad interest~\cite{wanjura_topological_2020,PhysRevLett.127.213601,PhysRevB.103.L241408,SciPostPhys.15.4.173,slim_optomechanical_2024}.
However, the topological origin of directional amplification in arbitrary dimensions is still undiscovered.

The directional amplification can be described by the response generating function.
The coefficients of the response generating function correspond to the response amplitudes of the lattice sites, so determining the spatial response is equivalent to extracting these coefficients.
In one dimension, the generating function is closely related to the Green’s function~\cite{PhysRevB.105.045122,PhysRevResearch.5.043073,PhysRevB.111.L060203,bai2026generating}.
In higher dimensions, however, the associated generating function becomes multivariate.
Although the generating function still encodes the full response, its inversion generally has no simple analytic expression.
In such cases, rather than inverting the generating function, one can study the asymptotic behavior of its coefficients, which determines the response at sites far from the drive.
A natural framework for this asymptotic treatment is Analytic Combinatorics in Several Variables (ACSV)~\cite{Pemantle2024,Melczer2021}, which yields the asymptotic expansion of these coefficients by the singularities of the generating function, \emph{i.e.}, the minimal critical points of its singular variety.
The logarithmic images of these points lie on the boundary of the amoeba~\cite{zbMATH00575960,mikhalkin_amoebas_2004} of the singular variety.
The concept of the amoeba has recently been used in non-Hermitian physics to formulate non-Bloch band theory in arbitrary dimensions and thereby to determine the spectrum~\cite{PhysRevX.14.021011}.

In this Letter, we trace the topological origin of directional amplification in non-Hermitian lattices of arbitrary dimensions to the topology of the amoeba complement.
Based on the response generating function and ACSV, we show that the asymptotic response is governed by the component of the amoeba complement that extends toward negative infinity, which we denote as $C_{-\infty}$.
Each component carries a distinct and discrete topological label named order~\cite{forsberg_laurent_2000}.
When the origin lies in the amoeba complement, comparing the order of the component
containing the origin with that of $C_{-\infty}$ yields a topological classification, where equal orders lead to exponential decay along all directions, and unequal orders lead to exponential amplification along some directions.
We use two- and three-dimensional non-Hermitian lattices for visual illustration, but our conclusions hold for arbitrary dimensions.

\runin{Response generating function}Consider an $n$-dimensional lattice Hamiltonian $H$, with the frequency-domain linear response to a local drive governed by
\begin{equation}\label{eq:EOM}
    \sum_{\mathbf{j'}}\left( \omega - H_{\mathbf{j,j'}} \right){\psi}_{\mathbf{j'}} = {d}_{\mathbf{j}}(\omega),
\end{equation}
where $\mathbf{j}=(j_1, j_2, \dots, j_n)$ labels the lattice site, ${\psi}_{\mathbf{j}}$ and ${d}_{\mathbf{j}}(\omega)$ are the response amplitude and the drive at site $\mathbf{j}$ in the frequency-domain.

We define the $n$-dimensional Z-transform of a response $\left\{ {\psi}_{\mathbf{j}} \right\}$ by
\begin{equation}
    \Psi(\mathbf{z})=\sum_{\mathbf{j}}{\psi}_{\mathbf{j}}\mathbf{z}^{\mathbf{j}},\quad \mathbf{z}^{\mathbf{j}} = {z_1}^{j_1}{z_2}^{j_2}\cdots{z_n}^{j_n},
\end{equation}
and call $\Psi(\mathbf{z})$ the response generating function.
In the translation-invariant bulk, $H_{\mathbf{j,j'}}=H_{\mathbf{j-j'}}  \equiv H_{\mathbf r}$, and the Z-transform of $\left\{ H_{\mathbf r} \right\}$ is
\begin{equation}
    H(\mathbf z)=\sum_{\mathbf r}H_{\mathbf r}\mathbf z^{\mathbf r}.
\end{equation}
Then Eq.~(\ref{eq:EOM}) is transformed to
\begin{equation}\label{eq:after-Z-transform}
    \left[ \omega - H(\mathbf{z}) \right]\Psi(\mathbf{z}) = D(\mathbf{z}) + b(\mathbf{z}),
\end{equation}
where $D(\mathbf{z})$ is the Z-transform of the local drive.
The term $b(\mathbf{z})$ is a boundary term determined by the boundary conditions, which encodes the truncation at the boundaries (see Sec.~S1 of the Supplemental Material~\cite{SM}).

Solving Eq.~(\ref{eq:after-Z-transform}) gives the generating function
\begin{equation}\label{eq:GeneratingFunc}
    \Psi(\mathbf{z}) = \frac{D(\mathbf{z}) + b(\mathbf{z}) }{\omega - H(\mathbf{z})}.
\end{equation}
The real-space response is the inverse Z-transform of $\Psi(\mathbf{z})$, \emph{i.e.}, the response at site $\mathbf{j}$ is the coefficient of $\mathbf{z}^{\mathbf{j}}$.
However, the multidimensional inverse Z-transform involves multiple contour integrals and generally has no simple analytic expression.
In order to extract the asymptotic response at sites far away from the drive, we employ ACSV, which gives the asymptotic expansion of these coefficients from the singularities of the generating function.

\runin{Asymptotic response}ACSV applies to generating functions that are rational functions of the form $P(\mathbf{z})/Q(\mathbf{z})$, where $P$ and $Q$ are complex polynomials.
For a finite-range translation-invariant lattice, the response generating function $\Psi(\mathbf{z})$ in Eq.~(\ref{eq:GeneratingFunc}) is of this form and falls within the scope of ACSV.

We denote the denominator of $\Psi(\mathbf{z})$ as $h(\mathbf{z})$.
The singular variety of $\Psi(\mathbf{z})$ is defined as
\begin{equation}\label{eq:SingularVariety}
    \mathcal{V}=\left\{ \mathbf{z}\in\left( \mathbb{C}^* \right)^n: h(\mathbf{z})=0 \right\},
\end{equation}
where $\mathbb{C}^*\equiv\left( \mathbb{C} \setminus \{0\} \right)$ is the set of complex numbers excluding 0.
We define the log-absolute-value mapping which maps a complex vector to the real space as
\begin{equation}
    \mathbf{u}=\log|\mathbf{z}|\equiv\left( \log|z_1|,\log|z_2|,\dots,\log|z_n| \right)\in\mathbb{R}^n.
\end{equation}
The amoeba is the log-absolute-value image of the singular variety:
\begin{equation}
    \mathcal{A}_h=\left\{ \log|\mathbf{z}|:\mathbf{z}\in\mathcal{V} \right\}\subset\mathbb{R}^n.
\end{equation}
The complement $\mathbb{R}^n\setminus\mathcal{A}_h$ consists of connected components $C$, each of which is convex~\cite{Itenberg2009}.
The log-absolute-value image of the domain of convergence of $\Psi(\mathbf{z})$ is the component of $\mathbb{R}^n\setminus\mathcal{A}_h$ that is unbounded toward $\mathbf{u}\to-\infty$ (\emph{i.e.}, $u_i\to-\infty$ for every $i$). We call this component $C_{-\infty}$ and its boundary $\partial C_{-\infty}$.

We consider the asymptotic response along the asymptotic direction $\hat{\mathbf{r}}=(r_1,r_2,..., r_n)\in \mathcal{R}_{\ge0}^n\equiv[0, \infty)^n$. We adopt the $L^{1}$ normalization, \emph{i.e.}, $\lVert\hat{\mathbf{r}}\rVert_1=\sum_{i=1}^{n}r_i=1$.
According to ACSV, a point $\mathbf{z}_c\in\mathcal{V}$ is named a critical point along the direction $\hat{\mathbf{r}}$ if there exists a nonzero complex number $\lambda$ such that
\begin{equation}\label{eq:CriticalEq}
    \nabla_{\log} h(\mathbf z_c)=\lambda\hat{\mathbf r},
\end{equation}
where $\nabla_{\log} h(\mathbf z)\equiv(z_1\partial_{z_1}h,\dots,z_n\partial_{z_n}h)$.
The asymptotic response along $\hat{\mathbf{r}}$ is governed by the minimal critical points $\mathbf{z}_{\min}$, \emph{i.e.}, the critical points whose log-absolute-value images lie on $\partial C_{-\infty}$~\cite{Pemantle2024}.

Under the standard smoothness and nondegeneracy conditions of ACSV, the asymptotic response contributed by a single minimal critical point $\mathbf{z}_{\min}$ takes the form
\begin{equation}\label{eq:ACSV_formula}
    \psi_{\mathbf{j}=\lfloor N\hat{\mathbf{r}} \rfloor}\sim A(\mathbf{z}_{\min},\hat{\mathbf r})\,N^{(1-n)/2} e^{\kappa_c(\hat{\mathbf r})N},
\end{equation}
where $\lfloor N\hat{\mathbf{r}} \rfloor$ denotes the componentwise floor of $N\hat{\mathbf{r}}$, $A$ is a coefficient determined by $\mathbf{z}_{\min}$ and $\hat{\mathbf r}$, and
\begin{equation}\label{eq:kappa_c}
    \kappa_c(\hat{\mathbf r}) = - \hat{\mathbf r} \cdot \log \mathbf{z}_{\min}
        = - \hat{\mathbf r} \cdot \left( \log |\mathbf{z}_{\min}| + \mathrm{i} \arg \mathbf{z}_{\min} \right).
\end{equation}
Its real part $\kappa(\hat{\mathbf r})=- \hat{\mathbf r} \cdot\log |\mathbf{z}_{\min}|$ is the exponential rate.
If there are several minimal critical points for a certain direction, their contributions will add up and interfere.
If two or more of them coalesce, the power-law exponent changes.
A detailed derivation of Eq.~(\ref{eq:ACSV_formula}) for a two-dimensional case, together with an illustration of the interference and coalescence of multiple minimal critical points, is provided as an example in Sec.~S2 of the Supplemental Material~\cite{SM}.
The response shows an exponential amplification behavior along $\hat{\mathbf r}$ for $\kappa(\hat{\mathbf r})>0$, and an exponential decay behavior for $\kappa(\hat{\mathbf r})<0$, while for $\kappa(\hat{\mathbf r})=0$, the response follows a power law.

\begin{figure}[t]
    \centering
    \begin{center}
        \includegraphics[width=0.6\linewidth]{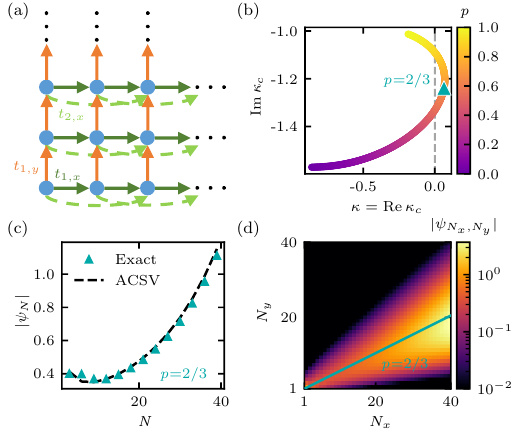}
    \end{center}
    \caption{(a) Schematic of a two-dimensional unidirectional model with open boundary conditions. The blue circles denote lattice sites, each with a uniform on-site loss $\gamma$. $t_{1,x}$ and $t_{2,x}$ are the unidirectional nearest-neighbor (NN) and next-nearest-neighbor (NNN) hopping amplitudes along the $x$ direction, and $t_{1,y}$ is the unidirectional NN hopping amplitude along the $y$ direction. In (b)-(d), we use $t_{1,x}=t_{2,x}=t_{1,y}=0.4$, $\gamma=1$. A local drive of amplitude $d_0=1$ is applied to the site $(1,1)$ of the lattice with a size of $40\times40$. (b) Trajectory of the complex exponential rate $\kappa_c$ of the minimal critical point as a function of the asymptotic direction $\hat{\mathbf{r}}=(p,1-p)$. The real part $\kappa=\mathrm{Re}\,\kappa_c$ is the exponential rate. $\kappa>0$ and $\kappa<0$ correspond to amplification and decay, respectively. (c) ACSV asymptotic form (dashed curve) and exact response $|\psi_N|$ (triangles) for $p=2/3$. (d) Two-dimensional response $|\psi_{N_x,N_y}|$. The cross section along the direction of $p=2/3$ corresponds to (c).}
    \label{fig:ACSV}
\end{figure}

To illustrate the framework, we now consider a two-dimensional unidirectional model with open boundary conditions shown in Fig.~\ref{fig:ACSV}(a) as an example.
In $z$-space, its Hamiltonian is
\begin{equation}
    H(z_x, z_y) = \omega_0 - \mathrm{i}\gamma + t_{1,x}{z_x} + t_{2,x}{z_x}^2 + t_{1,y}{z_y},
\end{equation}
where $\omega_0$ is the on-site frequency, $\gamma$ is the on-site loss and the $t$s are unidirectional hopping amplitudes.
For simplicity, we consider a local drive of amplitude $d_0$ and frequency $\omega=\omega_0$ at site $(1,1)$. Then, the response generating function becomes
\begin{equation}
    \Psi(z_x, z_y) = \frac{d_0 z_x z_y}{h(z_x, z_y)},
\end{equation}
where $h(z_x, z_y)=\omega_0-H(z_x, z_y)$.
Because the hopping is unidirectional, a site cannot influence any upstream sites.
For the response inside the lattice, the finite open lattice is therefore equivalent to a semi-infinite lattice, thus the boundary term $b(z_x,z_y)$ does not affect the response (see Sec.~S1B of the Supplemental Material~\cite{SM}).
Denoting the asymptotic direction as $\hat{\mathbf{r}}=(p,1-p)$ (due to the $L^1$ normalization), we find the critical points using $h(\mathbf{z})=0$ and Eq.~(\ref{eq:CriticalEq}) for different $p\in[0,1]$.
Then, the minimal critical points are obtained by retaining only the critical points whose log-absolute-value images lie on $\partial C_{-\infty}$.
The resulting trajectory of $\kappa_c$ in Eq.~(\ref{eq:kappa_c}) is drawn in Fig.~\ref{fig:ACSV}(b).
As an example, we compare the asymptotic response in Eq.~(\ref{eq:ACSV_formula}) with the exact result in Fig.~\ref{fig:ACSV}(c) for the direction of $p=2/3$, which corresponds to a cross section of the two-dimensional response along the given direction [see Fig.~\ref{fig:ACSV}(d)].
The asymptotic and exact results agree well for large $N$.

\begin{figure}[t]
    \centering
    \includegraphics[width=\linewidth]{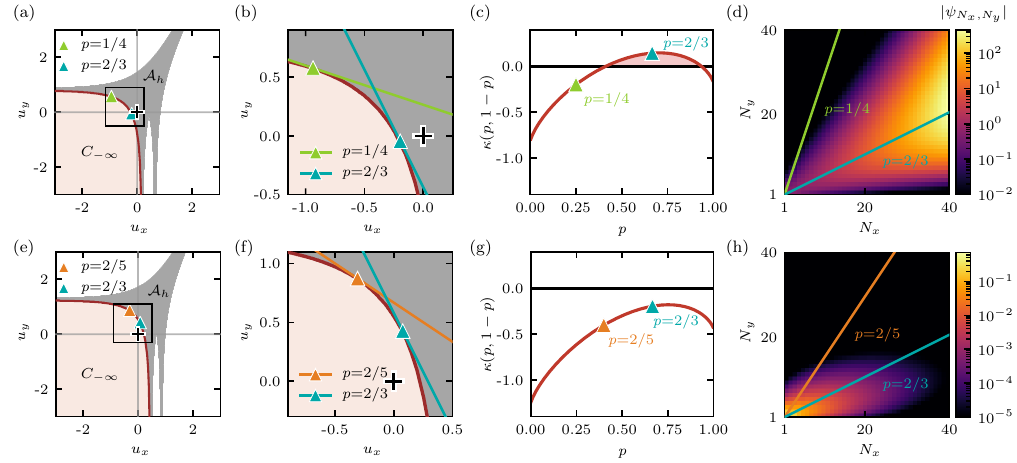}
    \caption{Amoeba geometry and directional amplification for two parameter sets. Top row (a)-(d): directional amplification. Bottom row (e)-(h): no directional amplification. Common parameters: $t_{1,x}=t_{2,x}=t_{1,y}=0.4$ and a local drive of amplitude $d_0=1$ is applied to the site $(1,1)$ of the lattice with a size of $40\times40$. Only the on-site loss differs, with $\gamma=0.9$ in (a)-(d) and $\gamma=1.4$ in (e)-(h). (a),(e) Amoeba $\mathcal{A}_h$ (gray) and the component $C_{-\infty}$ (pink; lower-left in this 2D view). The origin (black cross) lies outside $C_{-\infty}$ in (a) and inside it in (e). Triangles mark the logarithmic images of the minimal critical points for different asymptotic directions, which lie on the boundary of $C_{-\infty}$. (b),(f) Magnified view of the amoeba boundary and the supporting hyperplanes (solid lines) through different asymptotic directions. The relative position of the origin and $C_{-\infty}$ with respect to the supporting hyperplane determines the sign of the exponential amplification rate $\kappa$. Opposite sides of the supporting hyperplane correspond to amplification ($\kappa>0$), whereas the same side corresponds to decay ($\kappa<0$). (c),(g) Exponential amplification rate $\kappa$ versus the asymptotic direction $p$. $\kappa>0$ for some directions in (c), whereas $\kappa<0$ for all directions in (g). (d),(h) Two-dimensional response $|\psi_{N_x,N_y}|$ on a logarithmic color scale: amplified along some directions in (d) and decaying along every direction in (h).}
    \label{fig:amoeba}
\end{figure}

\begin{figure}[t]
    \centering
    \includegraphics[width=0.6\linewidth]{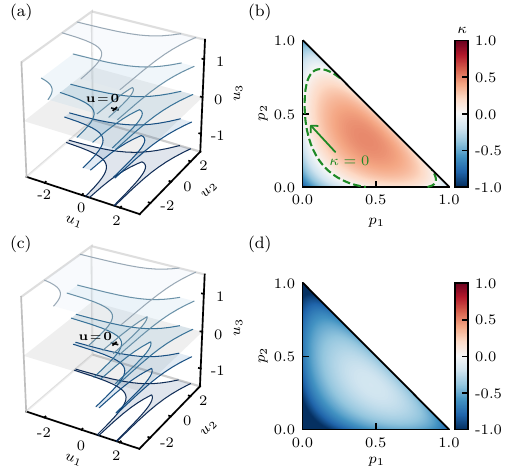}
    \caption{Amoeba geometry and directional amplification in three dimensions for two parameter sets. Top row (a),(b): $\gamma=1.5$; bottom row (c),(d): $\gamma=3$. Common parameters: $t_{1,x}=1.0$, $t_{2,x}=0.2$, $t_{1,y}=0.8$, and $t_{1,z}=0.6$. (a),(c) Slices of the three-dimensional amoeba $\mathcal{A}_h$ at different heights $u_3$, shown with a blue gradient that darkens from top to bottom. The gray plane marks the plane $u_3=0$, which contains the origin $\mathbf{u}=\mathbf{0}$ (black cross). In (a), the origin lies outside $C_{-\infty}$ (the component of the amoeba complement that is unbounded toward $\mathbf{u}\to-\infty$), whereas in (c) it lies inside $C_{-\infty}$. (b),(d) Heat maps of the exponential amplification rate $\kappa(p_1,p_2, 1-p_1-p_2)$. Only non-negative directions ($p_1,p_2\ge0$ and $p_1+p_2\le1$) are considered.
    Consistent with the classification, the origin lying outside $C_{-\infty}$ in (a) corresponds to $\kappa>0$ for some non-negative directions in (b), whereas the origin lying inside $C_{-\infty}$ in (c) corresponds to $\kappa<0$ for all non-negative directions in (d).
    }
    \label{fig:amoeba_3d}
\end{figure}

\runin{Topological origin of directional amplification}As introduced above, the log-absolute-value images of the minimal critical points $\mathbf{z}_{\min}$ lie on \(\partial C_{-\infty}\), where \(C_{-\infty}\) is the component of \(\mathbb{R}^n\setminus\mathcal{A}_h\) unbounded toward \(\mathbf{u}\to-\infty\) [see Fig.~\ref{fig:amoeba}(a) and (e)].
Because \(C_{-\infty}\) is convex, we can construct a supporting hyperplane through $\mathbf{u}_m=\log|\mathbf{z}_{\min}|$ to formulate a criterion for directional amplification.
If the boundary \(\partial C_{-\infty}\) is smooth at $\mathbf{u}_m$, the hyperplane is simply the tangent hyperplane and its normal vector is the asymptotic direction $\hat{\mathbf{r}}$.
Even if the boundary is not smooth, there still exists a supporting hyperplane that passes through this point and leaves the entire set $C_{-\infty}$ in one of its closed half-spaces.
Since this hyperplane touches $\partial C_{-\infty}$ at $\mathbf{u}_m$ and leaves all of $C_{-\infty}$ on one side, the position of the origin relative to the hyperplane determines the sign of the exponential rate $\kappa(\hat{\mathbf{r}})$.
In the following, we use the two- and three-dimensional cases in Fig.~\ref{fig:amoeba} and Fig.~\ref{fig:amoeba_3d} for visual demonstration, but the theoretical framework applies to arbitrary dimensions.

For a non-negative direction $\hat{\mathbf r}\in \mathcal{R}_{\ge0}^n$, the supporting hyperplane of $C_{-\infty}$ touches $\partial C_{-\infty}$ at $\mathbf{u}_m$.
This supporting hyperplane is determined by
$\hat{\mathbf r}\cdot\mathbf{u}+\kappa(\hat{\mathbf r})=0$, where $\kappa(\hat{\mathbf r})=-\hat{\mathbf r}\cdot\mathbf{u}_m$ is the exponential rate along $\hat{\mathbf r}$.
Thus, the parallel hyperplane through the origin, $\hat{\mathbf r}\cdot\mathbf{u}=0$, is the $\kappa=0$ reference.
The response is amplified (with $\kappa(\hat{\mathbf r})>0$) when the origin and \(C_{-\infty}\) lie on different sides of the supporting hyperplane, and decays (with $\kappa(\hat{\mathbf r})<0$) when they lie on the same side.
When the origin lies on the hyperplane, the response follows a power law (with $\kappa(\hat{\mathbf r})=0$).
For example, in Fig.~\ref{fig:amoeba}(b), the hyperplane separates the origin from \(C_{-\infty}\) for the direction of $p=2/3$, whereas it leaves them on the same side for the direction of $p=1/4$, resulting in different signs of $\kappa(\hat{\mathbf r})=\kappa(p,1-p)$ and amplification behaviors as shown in Figs.~\ref{fig:amoeba}(c) and (d).

For the above case in Fig.~\ref{fig:amoeba}(a), we can see that the origin ($\mathbf{u}=0$) lies outside the closure of \(C_{-\infty}\).
Because \(C_{-\infty}\) is convex, there exist non-negative directions for which the supporting hyperplane separates the origin from \(C_{-\infty}\).
For these directions $\kappa(\hat{\mathbf r})>0$ and the response is exponentially amplified.
If the origin lies on \(\partial C_{-\infty}\) instead, then $\kappa(\hat{\mathbf r})\le0$. Only the normal direction of the supporting hyperplane at the origin has $\kappa(\hat{\mathbf r})=0$ with a response following a power law along this direction, while all other directions result in $\kappa(\hat{\mathbf r})<0$ with exponentially decaying responses.
If the origin lies within \(C_{-\infty}\) as shown in Fig.~\ref{fig:amoeba}(e), it is on the same side as \(C_{-\infty}\) with regard to the supporting hyperplane for all non-negative directions, resulting in $\kappa(\hat{\mathbf r})<0$ [see Figs.~\ref{fig:amoeba}(f) and (g)]. Therefore, the response decays exponentially in all directions [see Fig.~\ref{fig:amoeba}(h)].

To conclude, generally,
\begin{equation}
    \mathbf 0\in C_{-\infty}\cup \partial C_{-\infty}
    \quad\Longleftrightarrow\quad
    \kappa(\hat{\mathbf r})\le 0\ \ \forall\hat{\mathbf r}\in\mathcal{R}_{\ge0}^n,
\end{equation}
Otherwise, there must exist some directions with $\kappa(\hat{\mathbf r})>0$. The detailed proof of this result is given in Sec.~S3A of the Supplemental Material~\cite{SM}.
Thus, the topological property of $C_{-\infty}$ provides the topological origin of the directional amplification behavior.

To further classify the topology, we can use the Ronkin function~\cite{Ronkin1974}
\begin{equation}
    R_h(\mathbf u)=\int_{[0,2\pi]^n}\log|h(e^{\mathbf u+\mathrm i\boldsymbol\theta})|
    \left(\frac{d\theta}{2\pi}\right)^n,
\end{equation}
where \(\mathbf u\in\mathbb R^n\) and $\boldsymbol\theta\in[0,2\pi]^n$. The gradient \(\nabla R_h(\mathbf u)\) is constant on each connected component of \(\mathbb R^n\setminus\mathcal A_h\), and more importantly, distinct components have distinct gradients~\cite{forsberg_laurent_2000}. Further details are given in Sec.~S3B of the Supplemental Material~\cite{SM}. Therefore, \(\nabla R_h(\mathbf u)\) can be used to define the order of the component $C$, \emph{i.e.}, \(\mathrm{ord}(C)\equiv\nabla R_h(\mathbf u)\) for any \(\mathbf u\in C\), which provides a discrete topological label for each component.

For \(\mathbf 0\in C_{-\infty}\), as the order labels the component uniquely,
\begin{equation}
    \mathrm{ord}(C_{-\infty})=\nabla R_h(\mathbf 0).
\end{equation}
In this case, the response decays exponentially along all non-negative directions.

For \(\mathbf 0\notin C_{-\infty}\), if \(\mathbf 0\notin\mathcal A_h\),
\begin{equation}
    \mathrm{ord}(C_{-\infty})\neq\nabla R_h(\mathbf 0).
\end{equation}
In this case, the response can be exponentially amplified along some non-negative directions.

On the other hand, if \(\mathbf 0\in\mathcal A_h\), the order of $\mathcal A_h$ is undefined. More specifically, we can divide this into two different cases. If \(\mathbf 0\in\partial C_{-\infty}\), the response follows a power law along the normal direction of the supporting hyperplane through the origin and decays exponentially along all other directions. If \(\mathbf 0\in\mathcal A_h\setminus\partial C_{-\infty}\), the response can be exponentially amplified along some non-negative directions.

For further demonstration, we consider a three-dimensional unidirectional lattice with
\begin{equation}
    H_{\mathrm{3D}}(\mathbf{z}) = \omega_0 - \mathrm{i}\gamma + t_{1,x}z_1 + t_{2,x}{z_1}^2 + t_{1,y}z_2 + t_{1,z}z_3.
\end{equation}
For each non-negative direction $\hat{\mathbf r}=(p_1,p_2,1-p_1-p_2)$, we solve the critical equations and find the minimal critical points on $\partial C_{-\infty}$.
Figs.~\ref{fig:amoeba_3d}(a) and (c) show slices of the three-dimensional amoeba for $\gamma=1.5$ and $\gamma=3$, and Figs.~\ref{fig:amoeba_3d}(b) and (d) show the corresponding exponential rates $\kappa(p_1,p_2, 1-p_1-p_2)$.
For $\gamma=1.5$, the origin lies outside $C_{-\infty}$, and $\kappa$ can be larger than 0 for some directions, corresponding to directional amplification.
For $\gamma=3$, the origin lies inside $C_{-\infty}$, and $\kappa<0$ for all directions, corresponding to exponential decay.
These results for the three-dimensional case are also consistent with the general conclusions for arbitrary dimensions.

\runin{Discussions and conclusions}In conclusion, we have identified the topological origin of directional amplification in driven non-Hermitian lattices of arbitrary dimensions, based on the investigation of the response generating function and the amoeba of its singular variety.
Using ACSV, we have shown that the asymptotic response is governed by the minimal critical points on \(\partial C_{-\infty}\), with the exponential rate determined by the position of the origin relative to the corresponding supporting hyperplane.
Because the orders label the components uniquely, we obtain a topological criterion that for the origin in the complement of the amoeba, the response decays exponentially along all directions if and only if the order of the component containing the origin is equal to that of \(C_{-\infty}\).
Our theory can be applied to various platforms including photonic, electrical, and acoustic systems~\cite{Wang:23,sahin_topolectrical_2025,huang_acoustic_2024}.

During the writing up of this letter, we noticed a related paper on arXiv~\cite{wanjura2025unifying}, which is based on the singular value decomposition and connects the band topology of a non-Hermitian lattice to directional amplification in one and two dimensions.
As its authors noted in their paper, the systematic exploration of higher-dimensional systems is currently still technically demanding.
However, our theory is based on ACSV and amoeba, and is not limited to lower dimensions but rather applies to arbitrary dimensions.

\runin{Acknowledgments}This research was supported by the National Natural Science Foundation of China (Grants No.~11974031), Natural Science Foundation of Jiangsu Province (BK20232002), and Innovation Program for Quantum Science and Technology (2021ZD0301702).

\runin{Data availability}All source data that support the plots within this paper and other findings of this study are available from the corresponding authors upon reasonable request.

\FloatBarrier
\printbibliography
\end{document}


\begin{center}
{\Large Supplemental Material: Topological Origin of Directional Amplification in Arbitrary-Dimensional Non-Hermitian Lattices\par}
\vspace{0.5em}
{\normalsize Zhenxiang Wu\textsuperscript{1}, Linfeng Zhang\textsuperscript{1,2,3}, and Haitan Xu\textsuperscript{1,2,3,*}\par}
\vspace{0.35em}
\begin{minipage}{0.95\textwidth}
\centering\small
\textsuperscript{1}National Laboratory of Solid State Microstructures, School of Advanced Manufacturing Engineering, Nanjing University, Suzhou 215163, China\\
\textsuperscript{2}Jiangsu Key Laboratory of Quantum Information Science and Technology, Nanjing University, Suzhou 215163, China\\
\textsuperscript{3}Shishan Laboratory, Nanjing University, Suzhou 215163, China\\[0.25em]
\textsuperscript{*}Corresponding author: \texttt{haitanxu@nju.edu.cn}
\end{minipage}
\end{center}

\section{Z-transform and boundary term}

\subsection{General formulation}

We start from the frequency-domain equation of motion in the main text [Eq.~(1)],
\begin{equation}
    \sum_{\mathbf j'\in\mathcal B}\left[\omega I-H\right]_{\mathbf j,\mathbf j'}\psi_{\mathbf j'}=d_{\mathbf j},
    \label{eq:SM_EOM}
\end{equation}
where \(\mathcal B=\{\mathbf j:1\le j_\alpha\le L_\alpha,\ \alpha=1,\dots,n\}\) is the finite box used in the main text.
For an infinite lattice, \(\mathcal B=\mathbb Z^n\).
In the bulk, translation invariance implies \(H_{\mathbf j,\mathbf j'}=H_{\mathbf j-\mathbf j'}\), so the same recurrence holds at every bulk site.
For a multi-index \(\mathbf j=(j_1,\dots,j_n)\) and \(\mathbf z=(z_1,\dots,z_n)\), we use \(\mathbf z^{\mathbf j} = {z_1}^{j_1}{z_2}^{j_2}\cdots{z_n}^{j_n}\) and define the \(Z\)-transform
\begin{equation}
    \Psi(\mathbf z)=\sum_{\mathbf j\in\mathcal B}\psi_{\mathbf j}\mathbf z^{\mathbf j},
    \qquad
    D(\mathbf z)=\sum_{\mathbf j\in\mathcal B}d_{\mathbf j}\mathbf z^{\mathbf j},
\end{equation}
consistent with Eq.~(2) of the main text, together with $z$-space Hamiltonian
\begin{equation}
    H(\mathbf z)=\sum_{\mathbf r}H_{\mathbf r}\mathbf z^{\mathbf r},
    \qquad
    H_{\mathbf r}\equiv H_{\mathbf j-\mathbf j'}.
\end{equation}

To obtain the equation in $z$-space, we multiply Eq.~\eqref{eq:SM_EOM} by \(\mathbf z^{\mathbf j}\) and sum over \(\mathbf j\in\mathcal B\).
The interior terms combine into \([\omega I-H(\mathbf z)]\Psi(\mathbf z)\).
Because the finite box is obtained by truncating a translation-invariant bulk model, the terms with \(\mathbf j'+\mathbf r\notin\mathcal B\) are absent from the finite sum.
Moving these missing contributions to the right-hand side gives
\begin{equation}
    \left[\omega I-H(\mathbf z)\right]\Psi(\mathbf z)=D(\mathbf z)+b(\mathbf z),
    \label{eq:SM_Zeq_general}
\end{equation}
where the boundary term $b(\mathbf z)$ collects exactly the hopping processes that would leave the finite box. $b(\mathbf z)$ is generally a Laurent polynomial supported at exponents outside \(\mathcal B\).
Although \(b(\mathbf z)\) appears on the right-hand side of Eq.~\eqref{eq:SM_Zeq_general}, it is not an independent external source: it is determined by the boundary values of \(\psi\) and enforces the open-boundary truncation.
For an infinite lattice, there are no such terms, so \(b(\mathbf z)=0\).
For a finite lattice, \(b(\mathbf z)\) is generally nonzero and encodes the truncation at the boundaries.
Therefore, it should be retained in a finite-lattice calculation in general.
Whether \(b(\mathbf z)\) affects the response depends on the model.
For the unidirectional model considered in the main text, the boundary term does not contribute to the response, as we explain below.
Consequently, it can be dropped when evaluating the system's response.

\subsection{Two-dimensional example}
In this subsection, we use the two-dimensional model introduced in the main text to illustrate the Z-transform procedure and to show explicitly how the boundary term arises.
We take the open boundary conditions (OBC) and the corresponding real-space Hamiltonian is
\begin{equation}
    H=\sum_{l=1}^{L_x}\sum_{m=1}^{L_y}\left[
    (\omega_0-\mathrm{i}\gamma)c_{l,m}^{\dagger}c_{l,m}
    +t_{1,x}c_{l+1,m}^{\dagger}c_{l,m}
    +t_{2,x}c_{l+2,m}^{\dagger}c_{l,m}
    +t_{1,y}c_{l,m+1}^{\dagger}c_{l,m} \right],
    \label{eq:SM_Hreal}
\end{equation}
where \(c_{l,m}\) is the annihilation operator at site \((l,m)\).
At frequency \(\omega=\omega_0\), the equation of motion [Eq.~\eqref{eq:SM_EOM}] in the bulk gives the recurrence
\begin{equation}
    \mathrm{i}\gamma\,\psi_{l,m}-t_{1,x}\psi_{l-1,m}-t_{2,x}\psi_{l-2,m}-t_{1,y}\psi_{l,m-1}=d_{l,m}.
    \label{eq:SM_real}
\end{equation}
The OBC are imposed by the convention
\begin{equation}
    \psi_{l,m}=0
    \qquad\text{for}\qquad
    l\le0\ \text{or}\ m\le0,
\end{equation}
so that Eq.~\eqref{eq:SM_real} holds for all \(1\le l\le L_x\), \(1\le m\le L_y\) and describes the open-boundary driven response of every lattice site.
At the upper edges \(l=L_x\) and \(m=L_y\), there is no hopping from \(l=L_x+1\) or \(m=L_y+1\) due to the unidirectionality of the model.

Multiplying the real-space recurrence [Eq.~\eqref{eq:SM_real}] by ${z_x}^l{z_y}^m$ and summing over the finite lattice, we obtain the shifted terms that generate the bulk contribution together with boundary contributions localized at the upper edges. To express these boundary contributions compactly, we introduce the partial Z-transforms
\begin{equation}
    \Psi_l(z_y)=\sum_{m=1}^{L_y}\psi_{l,m}{z_y}^m,
    \qquad
    \Psi_m(z_x)=\sum_{l=1}^{L_x}\psi_{l,m}{z_x}^l.
    \label{eq:SM_partial}
\end{equation}
With these definitions, the shifted terms become
\begin{align}
    \sum_{l,m}\psi_{l-1,m}{z_x}^l {z_y}^m
    &=z_x\Psi(z_x,z_y)-{z_x}^{L_x+1}\Psi_{L_x}(z_y),
    \label{eq:SM_shift1}\\
    \sum_{l,m}\psi_{l-2,m}{z_x}^l {z_y}^m
    &={z_x}^2\Psi(z_x,z_y)-{z_x}^{L_x+2}\Psi_{L_x}(z_y)
      -{z_x}^{L_x+1}\Psi_{L_x-1}(z_y),
    \label{eq:SM_shift2}\\
    \sum_{l,m}\psi_{l,m-1}{z_x}^l {z_y}^m
    &=z_y\Psi(z_x,z_y)-{z_y}^{L_y+1}\Psi_{L_y}(z_x).
    \label{eq:SM_shift3}
\end{align}
Substituting these expressions into Eq.~\eqref{eq:SM_real} gives
\begin{equation}
    h(z_x,z_y)\Psi(z_x,z_y)=D(z_x,z_y)+b(z_x,z_y),
    \label{eq:SM_Zeq2D}
\end{equation}
where \(D(z_x,z_y)=\sum_{l,m}d_{l,m}{z_x}^l {z_y}^m\), and
\begin{align}
    h(z_x,z_y) &= \omega_0 - H(z_x,z_y) = \mathrm{i}\gamma - t_{1,x}z_x - t_{2,x}{z_x}^2 - t_{1,y}z_y, \label{eq:SM_h}\\
    b(z_x,z_y) &= -\left(t_{1,x}+t_{2,x}z_x\right){z_x}^{L_x+1}\Psi_{L_x}(z_y)
    -t_{2,x}{z_x}^{L_x+1}\Psi_{L_x-1}(z_y)-t_{1,y}{z_y}^{L_y+1}\Psi_{L_y}(z_x). \label{eq:SM_b2D}
\end{align}
The lower-boundary contributions vanish because $\psi_{l,m}=0$ outside the lower edges; the nonzero terms in Eq.~\eqref{eq:SM_b2D} come from the upper edges $l=L_x$ and $m=L_y$. For a drive of amplitude $d_0$ at site $(1,1)$, $D(z_x,z_y)=d_0 z_x z_y$.

Although the boundary term \eqref{eq:SM_b2D} is nonzero due to the truncation at the downstream edges, it does not affect the response at any site within the finite lattice.
The physical reason is the purely unidirectional nature of the coupling in this model: there are no backward-propagating channels.
The boundary term effectively acts as a fictitious source induced by the missing hoppings at the open edges.
Because propagation is strictly unidirectional, any response generated at these edges can only propagate further downstream (outside the finite box) and can never travel back to affect the upstream sites.
Equivalently, downstream sites cannot affect upstream ones.
Therefore, the response of the finite lattice is indistinguishable from that of a semi-infinite lattice.
We can therefore drop $b(z_x,z_y)$ when extracting the response of the finite site:
\begin{equation}
    \Psi(z_x,z_y)=\frac{D(z_x,z_y)}{h(z_x,z_y)}
    =\frac{d_0 z_x z_y}{h(z_x,z_y)}.
    \label{eq:SM_bulk2D}
\end{equation}
This is the generating function used in the main text.

\subsection{Three-dimensional example}

The three-dimensional model of the main text is the direct generalization of Eq.~\eqref{eq:SM_Hreal}. On a lattice of size \(L_1\times L_2\times L_3\) with open boundary conditions,
\begin{equation}
\begin{split}
H=\sum_{l=1}^{L_1}\sum_{m=1}^{L_2}\sum_{k=1}^{L_3}\Big[
&(\omega_0-\mathrm{i}\gamma)c_{l,m,k}^{\dagger}c_{l,m,k}
+t_{1,x}c_{l+1,m,k}^{\dagger}c_{l,m,k}
+t_{2,x}c_{l+2,m,k}^{\dagger}c_{l,m,k}\\
&+t_{1,y}c_{l,m+1,k}^{\dagger}c_{l,m,k}
+t_{1,z}c_{l,m,k+1}^{\dagger}c_{l,m,k}\Big],
\end{split}
\label{eq:SM_Hreal3D}
\end{equation}
where \(c_{l,m,k}\) annihilates the site \((l,m,k)\) and all hoppings are again unidirectional. At \(\omega=\omega_0\), Eq.~\eqref{eq:SM_EOM} gives the bulk recurrence
\begin{equation}
\mathrm{i}\gamma\psi_{l,m,k}-t_{1,x}\psi_{l-1,m,k}-t_{2,x}\psi_{l-2,m,k}
-t_{1,y}\psi_{l,m-1,k}-t_{1,z}\psi_{l,m,k-1}=d_{l,m,k},
\label{eq:SM_real3D}
\end{equation}
with \(\psi_{l,m,k}=0\) whenever one of the indices is nonpositive.
Multiplying Eq.~\eqref{eq:SM_real3D} by \({z_1}^{l}{z_2}^{m}{z_3}^{k}\) and summing over the finite
lattice, the same summation by parts as in Eqs.~\eqref{eq:SM_shift1}-\eqref{eq:SM_shift3} produces the bulk term \(h(z_1,z_2,z_3)\Psi(z_1,z_2,z_3)\) together with a boundary term localized at the three upper faces \(l=L_1\), \(m=L_2\) and \(k=L_3\).
The latter carries the factors \({z_1}^{L_1+1}\), \({z_2}^{L_2+1}\) and \({z_3}^{L_3+1}\), respectively, \emph{i.e.}, one factor per direction, exactly as in Eq.~\eqref{eq:SM_b2D}.
Here
\begin{equation}
h(z_1,z_2,z_3)=\omega_0-H_{\mathrm{3D}}(z_1,z_2,z_3)
=\mathrm{i}\gamma-t_{1,x}z_1-t_{2,x}{z_1}^2-t_{1,y}z_2-t_{1,z}z_3,
\label{eq:SM_h3D}
\end{equation}
where \(H_{\mathrm{3D}}\) is given in Eq.~(18) of the main text.
Since the hopping parameters are unidirectional, the boundary term does not affect the response of any site inside the lattice and can be dropped. For a drive of amplitude \(d_0\) at \((1,1,1)\) the response generating function is thus
\begin{equation}
\Psi(z_1,z_2,z_3)=\frac{d_0 z_1 z_2 z_3}{h(z_1,z_2,z_3)},
\label{eq:SM_bulk3D}
\end{equation}
which is the three-dimensional counterpart of Eq.~\eqref{eq:SM_bulk2D}.
Each step in this procedure is not specific to three dimensions and the same steps apply with arbitrary dimensions, which allows the criterion of Sec.~S3 to be stated for general dimensions.

\section{ACSV in two dimensions}

In this section, we present a derivation of the two-dimensional critical-point equations of Analytic Combinatorics in Several Variables (ACSV) and the associated asymptotic formula as an intuitive example.
The general derivation in arbitrary dimensions can be found in the ACSV literature~\cite{Pemantle2024,Melczer2021}.
We also discuss two situations beyond the generic single-critical-point case: the interference of two minimal critical points with equal exponential rates, and the coalescence of two minimal critical points.
Both cases are illustrated using the two-dimensional model of Sec.~S1.

\subsection{Critical point equations and asymptotic formula}

Starting from the response generating function of Eq.~\eqref{eq:SM_bulk2D}, the response at site \((j,k)\) is the coefficient of \(\Psi(z_x,z_y)\) and can be extracted by the inverse Z-transform, which is a two-dimensional Cauchy integral:
\begin{equation}
    \psi_{j,k}=\frac{1}{(2\pi\mathrm{i})^2}\oint_{\mathcal C_x}\oint_{\mathcal C_y}
    \frac{D(z_x,z_y)}{h(z_x,z_y)}\,{z_x}^{-j-1}{z_y}^{-k-1}\,\mathrm dz_x\,\mathrm dz_y.
    \label{eq:SM_Cauchy2D}
\end{equation}

To evaluate this integral using ACSV, we assume that the singular variety \(\mathcal{V}=\{(z_x,z_y): h(z_x,z_y)=0\}\) is smooth at the relevant singularities. Specifically, for fixed \(z_y\), we assume the integrand possesses a simple pole at \(z_x=\phi(z_y)\), such that
\begin{equation}
    h(\phi(z_y),z_y)=0
    \qquad\text{and}\qquad
    \partial_{z_x}h(\phi(z_y),z_y)\neq0,
\end{equation}
where \(\phi(z_y)\) is an analytic function in a neighborhood of \(z_y\). We further assume that the numerator does not vanish at this pole, \emph{i.e.}, \(D(\phi(z_y),z_y)\neq0\). When the system possess multiple simple poles, their contributions to the residue simply add up. Thus, it suffices to evaluate the contribution from one simple pole and the full response is obtained by summing over all such poles.

Near the simple pole \(z_x=\phi(z_y)\), the denominator can be Taylor expanded as
\begin{equation}
    h(z_x, z_y) = \partial_{z_x}h\big|_{\phi(z_y)}\left[ z_x - \phi(z_y) \right] + \mathcal O\left(\left( z_x - \phi(z_y) \right)^2\right),
\end{equation}
which corresponds to a first-order pole to leading order. Applying the residue theorem to the \(z_x\)-contour integral for fixed \(z_y\), we obtain
\begin{equation}
\begin{split}
    \frac{1}{2\pi\mathrm{i}}\oint_{\mathcal{C}_x}\frac{D(z_x, z_y)}{h(z_x, z_y)}{z_x}^{-j-1}~\mathrm{d}z_x
    &= \frac{1}{2\pi\mathrm{i}}\oint_{\mathcal{C}_x}\frac{D(z_x, z_y)}{\partial_{z_x}h \cdot \left( z_x - \phi(z_y) \right)}{z_x}^{-j-1}~\mathrm{d}z_x\\
    &= \frac{D(\phi(z_y), z_y)}{\partial_{z_x}h(\phi(z_y), z_y)}\phi^{-j-1}(z_y).
\end{split}
\end{equation}

Substituting this result back into Eq.~\eqref{eq:SM_Cauchy2D}, the two-dimensional integral is reduced to a single contour integral over \(z_y\):
\begin{equation}
    \psi_{j,k} = \frac{1}{2\pi\mathrm{i}}\oint_{\mathcal{C}_y}\frac{D(\phi(z_y), z_y)}{\partial_{z_x}h(\phi(z_y), z_y)}\phi^{-j-1}(z_y){z_y}^{-k-1}~\mathrm{d}z_y.
    \label{eq:SM_1D_integral}
\end{equation}

To extract the asymptotic behavior of the response as the site indices go to infinity, we evaluate the contour integral using the saddle-point method. This method requires casting the integrand into the standard form \(f(z)e^{\lambda g(z)}\) with a sufficiently large parameter \(\lambda\).

Starting from Eq.~\eqref{eq:SM_1D_integral}, we factor out the coefficient part and exponentiate the powers of \(z_y\):
\begin{equation}
    A(z_y) = \frac{D(\phi(z_y), z_y)}{\partial_{z_x}h(\phi(z_y), z_y)} \frac{1}{\phi(z_y) z_y},
\end{equation}
such that the integral becomes
\begin{equation}
    \psi_{j,k} = \frac{1}{2\pi\mathrm{i}} \oint_{\mathcal{C}_y} A(z_y) e^{-j \log \phi(z_y) - k \log z_y} \,\mathrm dz_y.
\end{equation}
We focus on the asymptotic regime where \(j,k \to \infty\) along a fixed ray in the two-dimensional lattice. We introduce a large parameter \(N = j+k\) and define the asymptotic direction by the fractions \(p = j/N\) and \(q = k/N\), with \(p,q \ge 0\) and \(p+q=1\). The integral then takes the standard large-parameter form:
\begin{equation}
    \psi_{j,k}=\frac{1}{2\pi\mathrm{i}}\oint_{\mathcal C_y}
    A(z_y)\,e^{N\Phi(z_y)}\,\mathrm dz_y,
    \label{eq:SM_1Dint}
\end{equation}
with the phase function given by
\begin{equation}
    \Phi(z_y)=-p\log\phi(z_y)-q\log z_y.
    \label{eq:SM_A_Phi}
\end{equation}
Here we have used \(j=\lfloor Np\rfloor\) and \(k=\lfloor Nq\rfloor\), and we take the limit \(N\to\infty\). In this limit, the leading contribution comes from the saddle points associated with the singular variety \(h=0\).

The saddle points \(z_y\) are determined by the condition \(\Phi'(z_y)=0\). Since \(\Phi(z_y)=-p\log\phi(z_y)-q\log z_y\), we have
\begin{equation}
    \Phi'(z_y) = -\frac{p\,\phi'(z_y)}{\phi(z_y)} - \frac{q}{z_y} = 0.
\end{equation}
Using \(h(\phi(z_y),z_y)=0\) and the implicit function theorem,
\begin{equation}
    \phi'(z_y)=-\frac{\partial_{z_y}h(\phi(z_y),z_y)}{\partial_{z_x}h(\phi(z_y),z_y)},
    \label{eq:SM_phiprime}
\end{equation}
so the saddle-point condition becomes
\begin{equation}
    \frac{p}{\phi(z_y)\,\partial_{z_x}h(\phi(z_y),z_y)}
    =
    \frac{q}{z_y\,\partial_{z_y}h(\phi(z_y),z_y)}.
    \label{eq:SM_crit_eq}
\end{equation}
Equivalently, at the critical point \(\mathbf z_c=(\phi(z_y),z_y)\), this condition can be restated as follows: there exists a nonzero complex number \(\lambda\) such that
\begin{equation}\label{eq:CriticalEq}
    \nabla_{\log} h(\mathbf z_c)=\lambda\hat{\mathbf r},
\end{equation}
where \(\nabla_{\log} h(\mathbf z)\equiv(z_x\partial_{z_x}h, z_y\partial_{z_y}h)\) and $\hat{\mathbf{r}}=\left( p,q \right)$.
Eq.~(\ref{eq:CriticalEq}) is the critical condition used in the main text. We assume that the saddle is nondegenerate, \emph{i.e.}, \(\Phi''(z_y)\neq0\).

It should be noted that the critical equation Eq.~(\ref{eq:CriticalEq}) can admit multiple solutions. However, not all critical points contribute to the leading asymptotic behavior of the positive-power coefficients. In the framework of ACSV, the asymptotic expansion is dominated only by the minimal critical point \(\mathbf z_{\min}=(z_{x,\min},z_{y,\min})\), which lies on the boundary of the domain of convergence.
This boundary corresponds to the boundary of the component of the amoeba complement that is unbounded towards $-\infty$, denoted as $C_{-\infty}$~\cite{Pemantle2024}.

Under these assumptions, the saddle-point evaluation of Eq.~\eqref{eq:SM_1Dint} around \(\mathbf z_{\min}\) gives
\begin{equation}
    \psi_{j,k}\sim A(\mathbf z_{\min},p,q)\,N^{-1/2}e^{\kappa_c(p,q)N},
    \qquad
    \kappa_c(p,q)=-p\log z_{x,\min}-q\log z_{y,\min},
    \label{eq:SM_2D_asym}
\end{equation}
where \(A\) is a nonuniversal prefactor depending on \(D\), \(\partial h\), and \(\Phi''\). In the numerical comparison of the main text, the exponent is fixed by ACSV, while \(A\) is treated as a fitting parameter. In \(n\) dimensions, the same argument gives a prefactor scaling as \(N^{(1-n)/2}\). When several minimal critical points are present, the full asymptotic response is the sum of their individual contributions. The interference is discussed in Sec.~S2B.

To verify this asymptotic prediction numerically, we use the ACSV result for the response along a fixed ray \(p:q=j:k\). Taking \(N=j+k\), the response takes the form
\begin{equation}
    \psi_{\lfloor Np\rfloor,\lfloor Nq\rfloor} \propto \frac{1}{\sqrt{N}} e^{\kappa_c(p,q)N}.
\end{equation}
Taking the absolute value and the logarithm on both sides yields
\begin{equation}
    \log |\psi_{\lfloor Np\rfloor,\lfloor Nq\rfloor}| = -\frac{1}{2}\log N + \kappa N+ C',
\end{equation}
where $\kappa=\mathrm{Re}[\kappa_c(p,q)]$ and \(C'\) is a constant independent of \(N\). Therefore, the quantity \(\log |\psi_{\lfloor Np\rfloor,\lfloor Nq\rfloor}| + \frac{1}{2}\log N\) is predicted to be a linear function of \(N\). By checking whether the numerically extracted response satisfies this linear behavior as \(N\) increases, we can verify the validity of the ACSV theory.

As an explicit example, for the model of Sec.~S1,
\begin{equation}
    h(z_x,z_y)=\mathrm{i}\gamma-t_{1,x}z_x-t_{2,x}{z_x}^2-t_{1,y}z_y.
    \label{eq:SM_h2D_ACSV}
\end{equation}
Eliminating $z_y$ from $h=0$ and Eq.~\eqref{eq:SM_crit_eq} gives
\begin{equation}
    (1+q)t_{2,x}{z_x}^2+t_{1,x}z_x-p\,\mathrm{i}\gamma=0,
    \label{eq:SM_quadratic}
\end{equation}
whose two roots are
\begin{equation}
    z_x^{\pm}=\frac{-t_{1,x}\pm\sqrt{{t_{1,x}}^2+4(1-q^2)t_{2,x}\,\mathrm{i}\gamma}}{2(1+q)t_{2,x}}.
    \label{eq:SM_zx_roots}
\end{equation}
The corresponding $z_y^{\pm}$ follow from $h=0$. The minimal critical point is obtained by choosing the critical root that lies on $\partial C_{-\infty}$.

\subsection{Coalescence and interference of the minimal critical points}

The derivation of Eq.~\eqref{eq:SM_2D_asym} assumes a nondegenerate saddle with \(\Phi''(z_y)\neq0\). This assumption fails when two critical points coalesce. To see this explicitly, note that the saddle-point condition \(\Phi'(z_y)=0\) is a rational equation, which can be written as
\begin{equation}
    \Phi'(z_y) = \frac{f(z_y)}{g(z_y)} = 0,
\end{equation}
where \(f\) and \(g\) are polynomials. When two critical points merge at \(z_0\), the numerator acquires a double root, \emph{i.e.}, \(f(z_y) = A(z_y-z_0)^2 + \mathcal{O}\left((z_y-z_0)^3\right)\). Since \(g(z_0)\neq0\), it follows directly that
\begin{equation}
    \Phi'(z_0)=0, \qquad \Phi''(z_0) = \frac{f'(z_0)}{g(z_0)} = 0.
    \label{eq:SM_degen_cond}
\end{equation}
Consequently, the phase function must be expanded to third order, and the leading contribution to the integral is no longer Gaussian. Instead of \(\int e^{-N\rho^2}\,\mathrm d\rho\), the dominant integral along the steepest descent path takes the cubic form
\begin{equation}
    \int e^{-N\rho^3}\,\mathrm d\rho.
    \label{eq:SM_cubic}
\end{equation}
Evaluating this integral yields
\begin{equation}
    \psi_{j,k}\sim C' N^{-1/3}e^{\kappa_c N},
    \label{eq:SM_N13}
\end{equation}
so the power law changes from \(N^{-1/2}\) to \(N^{-1/3}\). More generally, an \(M\)-fold degenerate critical point, where the first \(M\) derivatives of \(\Phi\) vanish and the \((M+1)\)-th is nonzero, gives a scaling of \(N^{-1/(M+1)}\).

\begin{figure}[tb]
    \includegraphics[width=\linewidth]{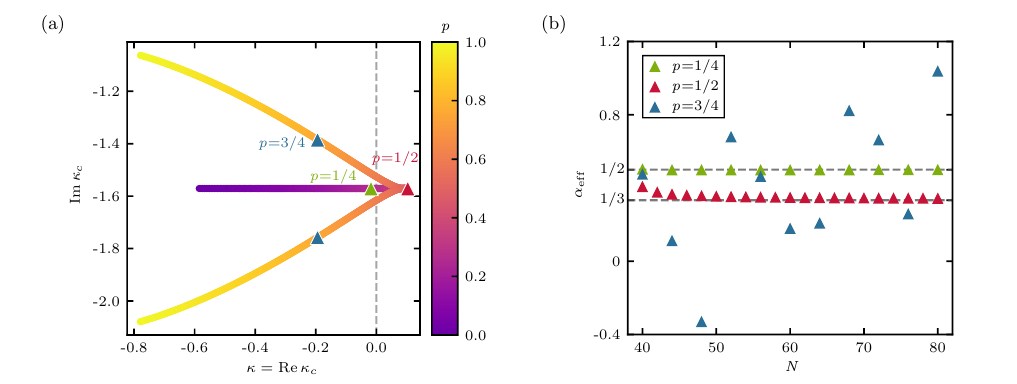}
    \caption{(a) Trajectory of the complex exponential rate $\kappa_c$ of the minimal critical points as a function of the asymptotic direction $p$. The parameters are $t_{1,x}=1$, $t_{2,x}=0.25\mathrm{i}$ (\emph{i.e.}, $\tau=0.25$), $t_{1,y}=0.7$, and $\gamma=4/3$, chosen such that $t_{1,x}^2=3\gamma\tau$ and hence $q_0=1/2$. Triangles mark three representative directions: for \(p=1/4\) (green), there is a single minimal critical point; for \(p=1/2\) (red), two minimal critical points coalesce into a twofold-degenerate critical point; and for \(p=3/4\) (blue), the response is governed by two minimal critical points with equal real parts of $\kappa_c$ and different imaginary parts, leading to interference. (b) Numerically extracted effective power-law exponent \(\alpha_{\mathrm{eff}}\) as a function of the asymptotic distance \(N=j+k\) for the three directions. Markers denote the extracted values for \(p=1/4\) (green), \(p=1/2\) (red), and \(p=3/4\) (blue). Horizontal dashed lines indicate the theoretical values \(\alpha=1/2\) (nondegenerate) and \(\alpha=1/3\) (twofold degenerate). To reduce the rapid oscillations induced by multi-saddle interference in the \(p=3/4\) case, a sliding-window finite difference with window size \(N_{\mathrm{win}}=40\) is used.}
    \label{fig:degeneracy}
\end{figure}

For the two-dimensional model of Sec.~S1, the discriminant of Eq.~\eqref{eq:SM_quadratic} determines when the two roots coincide. For real $t_{1,x}$ and $\gamma$, a coalescence requires a complex next-nearest-neighbor hopping; we take
\begin{equation}
    t_{2,x}=\mathrm{i}\tau,\qquad \tau\in\mathbb R.
    \label{eq:SM_complex_t2x}
\end{equation}
This choice is made only to illustrate the degeneracy, and the criterion itself does not depend on it. The discriminant of Eq.~\eqref{eq:SM_quadratic} is then
\begin{equation}
    \Delta={t_{1,x}}^2-4(1-q^2)\gamma\tau.
    \label{eq:SM_discriminant}
\end{equation}
The two critical points coalesce when $\Delta=0$, \emph{i.e.},
\begin{equation}
    1-{q_0}^2=\frac{{t_{1,x}}^2}{4\gamma\tau},
    \qquad 0<\frac{{t_{1,x}}^2}{4\gamma\tau}<1.
    \label{eq:SM_q0}
\end{equation}
For $q>q_0$, $\Delta>0$: the two roots are real and distinct, and the one lying on $\partial C_{-\infty}$ dominates. For $q<q_0$, $\Delta<0$: the roots are complex conjugates with equal modulus, so both contribute and interfere. At $q=q_0$, the roots coincide, $\Phi''=0$, and the power law is $N^{-1/3}$.

To verify the scaling laws predicted by ACSV and to clarify the effects of critical-point coalescence and multi-saddle interference on the asymptotic response, we plot the trajectory of the complex exponential rate \(\kappa_c\) of the minimal critical points and numerically extract the effective power-law exponent \(\alpha_{\mathrm{eff}}\). For the numerical example, we choose parameters such that the two critical points coalesce at \(p=1/2\), \emph{i.e.}, \(q_0=1/2\). According to Eq.~\eqref{eq:SM_q0}, this requires \(t_{1,x}^2=3\gamma\tau\) with \(t_{2,x}=\mathrm{i}\tau\). With this choice, for \(p<1/2\) (\(q>1/2\)) the two roots are real and only one of them is minimal, whereas for \(p>1/2\) (\(q<1/2\)) they are complex conjugates with equal modulus, so both are minimal and interfere; at \(p=1/2\) they coalesce. The parameters used in Fig.~\ref{fig:degeneracy} are given in its caption.

Figure~\ref{fig:degeneracy}(a) displays the trajectory of \(\kappa_c\) as the asymptotic direction \(p\) varies. The real part \(\mathrm{Re}\,\kappa_c\) governs the exponential growth or decay, while the imaginary part \(\mathrm{Im}\,\kappa_c\) determines the phase oscillation. The geometry of the trajectory determines the asymptotic behavior: (i) for \(p=1/4\), there is a single dominant branch (green triangle), corresponding to a single minimal critical point; (ii) for \(p=1/2\), the two branches merge at a single point (red triangle), corresponding to the coalescence of two critical points into a twofold-degenerate saddle; and (iii) for \(p=3/4\), the trajectory splits into two distinct branches (blue triangles), corresponding to two minimal critical points with equal real parts of \(\kappa_c\) and different imaginary parts, which contribute and interfere.

To extract the effective power-law exponent numerically, we write the asymptotic response along a fixed direction \(p:q=j:k\) as
\begin{equation}
    |\psi_{\lfloor Np\rfloor,\lfloor Nq\rfloor}| \sim A N^{-\alpha} e^{\kappa N},
\end{equation}
where \(A\) is a nonuniversal constant and \(\alpha\) depends on the degeneracy of the dominant critical point (e.g., \(\alpha=1/2\) for a simple saddle and \(\alpha=1/3\) for a twofold degenerate saddle). Taking the logarithm gives
\begin{equation}
    \ln |\psi_{\lfloor Np\rfloor,\lfloor Nq\rfloor}| = \ln A - \alpha \ln N + \kappa N.
\end{equation}
The exponent \(\alpha\) can be extracted from finite differences:
\begin{equation}
    \alpha = -\frac{\Delta \ln |\psi| - \kappa \Delta N}{\Delta \ln N}.
\end{equation}
Because the interference of multiple critical points induces rapid beat-like oscillations, adjacent-point differences would produce an \(\alpha\) with large fluctuations. We therefore use a sliding-window finite difference with span \(N_{\mathrm{win}}\) to obtain a coarse-grained estimate.

The extracted \(\alpha_{\mathrm{eff}}\) is shown in Fig.~\ref{fig:degeneracy}(b). For \(p=1/4\), the single minimal critical point gives the nondegenerate value \(\alpha_{\mathrm{eff}}=1/2\). For \(p=1/2\), \(\alpha_{\mathrm{eff}}\) converges asymptotically to \(1/3\), consistent with the twofold degeneracy identified in Fig.~\ref{fig:degeneracy}(a). For \(p=3/4\), the two minimal critical points have equal real parts and opposite imaginary parts of \(\kappa_c\), producing pronounced beat-like oscillations. Consequently, \(\alpha_{\mathrm{eff}}\) oscillates around the nondegenerate value \(1/2\) even after sliding-window averaging.

\section{Topological criterion for classifying the system response}

\subsection{Proof of the criterion}

We first recall two properties of the component \(C_{-\infty}\) and fix notation.
The component \(C_{-\infty}\) is the log-absolute-value image of the domain of convergence of \(\Psi(\mathbf z)\).
Since absolute convergence is preserved when any $|z_i|$ is decreased, decreasing any component of a point $\mathbf u=\log|\mathbf{z}|\in C_{-\infty}$ keeps it inside \(C_{-\infty}\):
\begin{equation}
    \mathbf u\in C_{-\infty},\quad v_i\le u_i, \forall i
    \quad\Longrightarrow\quad
    \mathbf v\in C_{-\infty}.
    \label{eq:SM_downward}
\end{equation}
Since \(C_{-\infty}\) is convex~\cite{Itenberg2009}, its closure \(\overline{C_{-\infty}}\) is also convex, and the property \eqref{eq:SM_downward} remains valid on \(\overline{C_{-\infty}}\), since decreasing a coordinate commutes with taking limits.

We now consider the three possible positions of the origin $\mathbf u=\mathbf 0$ relative to \(C_{-\infty}\).

\emph{Case 1: \(\mathbf 0\notin\overline{C_{-\infty}}\).}
Because \(\overline{C_{-\infty}}\) is a nonempty closed convex set, it contains a unique point \(\mathbf s\) closest to the origin.
Since \(\mathbf 0\notin\overline{C_{-\infty}}\), \(\mathbf s\neq\mathbf 0\).
We first show that every component of \(\mathbf s\) is nonpositive.
Suppose \(s_i>0\) for some \(i\).
Since \(\overline{C_{-\infty}}\) is downward closed, \(\mathbf s-\delta\mathbf e_i\in\overline{C_{-\infty}}\) for sufficiently small \(\delta>0\), where \(\mathbf e_i\) is the unit vector in the \(i\)-th coordinate.
The squared distance to the origin is
\begin{equation}
    |\mathbf s-\delta\mathbf e_i|^2
    =|\mathbf s|^2-2\delta s_i+\delta^2
    <|\mathbf s|^2
    \qquad(0<\delta<2s_i),
    \label{eq:SM_closer}
\end{equation}
which contradicts the choice of \(\mathbf s\) as the closest point.
Therefore \(s_i\le0\) for all \(i\).

To locate the supporting hyperplane, we use the fact that \(\mathbf s\) is the closest point to the origin.
For any \(\mathbf u\in\overline{C_{-\infty}}\) and any \(0\le t\le1\), convexity gives \((1-t)\mathbf s+t\mathbf u\in\overline{C_{-\infty}}\).
Since \(\mathbf s\) minimizes the distance to the origin,
\begin{equation}
    |(1-t)\mathbf s+t\mathbf u|^2\ge|\mathbf s|^2.
\end{equation}
Expanding the left-hand side for small \(t\) yields
\(2t\,\mathbf s\cdot(\mathbf u-\mathbf s)+t^2|\mathbf u-\mathbf s|^2\ge0\).
Dividing by \(t>0\) and letting \(t\to0^+\), we obtain
\begin{equation}
    \mathbf s\cdot(\mathbf u-\mathbf s)\ge0,
    \label{eq:SM_support}
\end{equation}
which means that $\mathbf s\cdot\mathbf u\ge|\mathbf s|^2$ for every \(\mathbf u\in\overline{C_{-\infty}}\).
Since \(\mathbf s\) is the point of \(\overline{C_{-\infty}}\) closest to the origin, it cannot lie in the open set \(C_{-\infty}\): a small ball around it would then also lie in \(C_{-\infty}\), and would contain points closer to the origin than \(\mathbf s\). Hence \(\mathbf s\in\partial C_{-\infty}\).
Geometrically, \eqref{eq:SM_support} says that the hyperplane \(\mathbf s\cdot\mathbf u=|\mathbf s|^2\) touches \(C_{-\infty}\) at this boundary point \(\mathbf s\in\partial C_{-\infty}\), with \(C_{-\infty}\) lying in the half-space \(\mathbf s\cdot\mathbf u\ge|\mathbf s|^2\) and the origin on the other side.

We now introduce the unit vector
\begin{equation}
    \hat{\mathbf r}\equiv-\frac{\mathbf s}{|\mathbf s|},
    \label{eq:SM_r}
\end{equation}
which is the outward normal of \(C_{-\infty}\) at \(\mathbf s\).
Since \(s_i\le0\) for all \(i\), its components satisfy \(\hat r_i\ge0\) for all \(i\), so \(\hat{\mathbf r}\in\mathcal R_{\ge0}^n\).
In terms of \(\hat{\mathbf r}\), Eq.~\eqref{eq:SM_support} gives
\begin{equation}
    \hat{\mathbf r}\cdot\mathbf u
    \le-|\mathbf s|<0,
    \quad
    \forall\mathbf u\in C_{-\infty}.
    \label{eq:SM_separation}
\end{equation}
Thus the hyperplane
\begin{equation}
    \hat{\mathbf r}\cdot\mathbf u+|\mathbf s|=0
    \label{eq:SM_hyperplane}
\end{equation}
supports \(C_{-\infty}\) at \(\mathbf s\) and strictly separates the origin from \(C_{-\infty}\): the set \(C_{-\infty}\) lies in the closed half-space \(\hat{\mathbf r}\cdot\mathbf u+|\mathbf s|\le0\), whereas the origin satisfies \(\hat{\mathbf r}\cdot\mathbf 0+|\mathbf s|=|\mathbf s|>0\).
The supporting point is \(\mathbf u_m=\mathbf s\), which is the log-absolute-value image of a minimal point on \(\partial C_{-\infty}\) for the direction \(\hat{\mathbf r}\).
With the convention \(\kappa(\hat{\mathbf r})=-\hat{\mathbf r}\cdot\mathbf u_m\) used in the main text, the corresponding rate is
\begin{equation}
    \kappa(\hat{\mathbf r})=|\mathbf s|>0,
\end{equation}
so the response is exponentially amplified along the non-negative direction \(\hat{\mathbf r}\).

\emph{Case 2: \(\mathbf 0\in C_{-\infty}\).}
Since \(C_{-\infty}\) is a connected component of \(\mathbb R^n\setminus\mathcal A_h\), it is open. Hence the origin is an interior point of \(C_{-\infty}\).
We consider a non-negative direction \(\hat{\mathbf r}\in\mathcal R_{\ge0}^n\).
As in the main text, let \(\mathbf u_m\in\partial C_{-\infty}\) be the point at which the supporting hyperplane of \(C_{-\infty}\) with outward normal \(\hat{\mathbf r}\) touches \(C_{-\infty}\).
The hyperplane is
\begin{equation}
    \hat{\mathbf r}\cdot\mathbf u+\kappa(\hat{\mathbf r})=0,
    \qquad
    \kappa(\hat{\mathbf r})=-\hat{\mathbf r}\cdot\mathbf u_m.
\end{equation}
Because \(\hat{\mathbf r}\) is the outward normal, the whole component lies on the side \(\hat{\mathbf r}\cdot\mathbf u\le\hat{\mathbf r}\cdot\mathbf u_m=-\kappa(\hat{\mathbf r})\), \emph{i.e.},
\begin{equation}
    \mathbf u\in C_{-\infty}
    \quad\Longrightarrow\quad
    \hat{\mathbf r}\cdot\mathbf u+\kappa(\hat{\mathbf r})\le0.
\end{equation}
Since \(\mathbf 0\in C_{-\infty}\), the origin lies in the same closed half-space:
\begin{equation}
    \hat{\mathbf r}\cdot\mathbf 0+\kappa(\hat{\mathbf r})=\kappa(\hat{\mathbf r})\le0.
\end{equation}
Moreover, if the origin lay on the supporting hyperplane, it would be a boundary point of \(C_{-\infty}\), since \(C_{-\infty}\) lies entirely in one closed half-space of this hyperplane.
This contradicts the fact that \(\mathbf 0\) is an interior point of \(C_{-\infty}\).
Therefore the inequality is strict,
\begin{equation}
    \kappa(\hat{\mathbf r})<0,
\end{equation}
and the response decays exponentially along \(\hat{\mathbf r}\).
Since the direction \(\hat{\mathbf r}\in\mathcal R_{\ge0}^n\) is arbitrary, the response decays exponentially along all non-negative directions.

\emph{Case 3: \(\mathbf 0\in\partial C_{-\infty}\).}
Again consider \(\hat{\mathbf r}\in\mathcal R_{\ge0}^n\) and let \(\mathbf u_m\in\partial C_{-\infty}\) be the supporting point with outward normal \(\hat{\mathbf r}\).
The same half-space argument gives
\begin{equation}
    \kappa(\hat{\mathbf r})\le0.
\end{equation}
If the supporting hyperplane does not pass through the origin, then the origin is strictly inside this half-space, so \(\kappa(\hat{\mathbf r})<0\).
If the supporting hyperplane passes through the origin, then
\begin{equation}
    \hat{\mathbf r}\cdot\mathbf 0+\kappa(\hat{\mathbf r})=0
    \quad\Longrightarrow\quad
    \kappa(\hat{\mathbf r})=0.
\end{equation}
Thus \(\kappa(\hat{\mathbf r})=0\) exactly for the normal direction of the supporting hyperplane that passes through the origin, and \(\kappa(\hat{\mathbf r})<0\) for all other directions.

These three cases are exhaustive and establish the classification used in the main text.

\subsection{Ronkin function and the order of the components of the amoeba complement}

For $\mathbf u\in\mathbb R^n\setminus\mathcal A_h$, the torus
$\{|z_j|=e^{u_j}:1\le j\le n\}$ does not intersect the singular variety $h=0$.
The Ronkin function is~\cite{Ronkin1974}
\begin{equation}
    R_h(\mathbf u)
    =\int_{[0,2\pi]^n}\log|h(e^{\mathbf u+i\boldsymbol\theta})|
    \prod_{k=1}^n\frac{d\theta_k}{2\pi}.
    \label{eq:SM_Ronkin}
\end{equation}
It is finite and smooth on each component of $\mathbb R^n\setminus\mathcal A_h$.
Differentiating with respect to $u_j$ and using
$z_j\partial_{z_j}h/h=-i\partial_{\theta_j}\log h$ gives
\begin{equation}
    \partial_{u_j}R_h(\mathbf u)
    =\int_{[0,2\pi]^n}\mathrm{Re}\left[\frac{z_j\partial_{z_j}h}{h}\right]
    \prod_{k=1}^n\frac{d\theta_k}{2\pi}
    =\int_{[0,2\pi]^{n-1}}w_j(\mathbf u;\{\theta_k\}_{k\ne j})
    \prod_{k\ne j}\frac{d\theta_k}{2\pi},
    \label{eq:SM_Ronkin_deriv}
\end{equation}
where
\begin{equation}
    w_j(\mathbf u;\{\theta_k\}_{k\ne j})
    =\frac{1}{2\pi\mathrm{i}}\int_0^{2\pi}d\theta_j\,
    \frac{\partial_{\theta_j}h}{h}
    \label{eq:SM_winding}
\end{equation}
is the winding number of $h$ along the $j$-th circle, with the other angles
$\theta_k$ ($k\ne j$) held fixed.
In the second equality of Eq.~\eqref{eq:SM_Ronkin_deriv}, the integral over
$\theta_j$ reduces to this winding number.
The remaining integral is over the $n-1$ angles $\theta_k$ with $k\ne j$.
Each $w_j$ is an integer.
As long as $\mathbf u$ stays inside a connected component of $\mathbb R^n\setminus\mathcal A_h$, $h$ never vanishes on the torus, so $w_j$ cannot change continuously; it is constant.
Consequently, $\nabla R_h$ is constant on the component,
\begin{equation}
    \nabla R_h(\mathbf u)=\boldsymbol\nu,
    \qquad
    \boldsymbol\nu\in\mathbb Z^n,
    \label{eq:SM_order}
\end{equation}
and $R_h(\mathbf u)=\boldsymbol\nu\cdot\mathbf u+c$ on that component. We call $\boldsymbol\nu$ the order of the component, denoted $\mathrm{ord}(C)$~\cite{forsberg_laurent_2000}.

The order is a topological invariant: it is integer-valued and constant on each component, and it can change only when $\mathbf u$ crosses the amoeba. Moreover, distinct components have distinct orders~\cite{forsberg_laurent_2000}. Thus each component carries a unique integer-vector label.

For $\mathbf 0\notin\mathcal A_h$, the origin lies in a component, and $\nabla R_h(\mathbf 0)$ is its order. Since the order map is injective,
\begin{equation}
    \nabla R_h(\mathbf 0)=\mathrm{ord}(C_{-\infty})
    \quad\Longleftrightarrow\quad
    \mathbf 0\in C_{-\infty}.
    \label{eq:SM_order_criterion}
\end{equation}
Combining this with the result of Sec.~S3A, we obtain the topological classification used in the main text: the response decays exponentially along all non-negative directions if and only if $\nabla R_h(\mathbf 0)=\mathrm{ord}(C_{-\infty})$. If $\mathbf 0\in\mathcal A_h$, the order is not defined, which is the critical case discussed in the main text.

\printbibliography